\documentclass[preprint2]{aastex}
\usepackage{spr-astr-addons}
\usepackage{graphicx}
\newcommand{\gtapprox}{\raisebox{-0.5ex}{$\,\stackrel{>}{\scriptstyle
\sim}\,$}}
\newcommand{\ltapprox}{\raisebox{-0.5ex}{$\,\stackrel{<}{\scriptstyle
\sim}\,$}}
\newcommand{\msolar}{M_\odot}

\begin{document}
\title{X-ray observations of ultraluminous X-ray sources}

\shorttitle{X-ray observations of ULXs}        
\shortauthors{T.\,P. Roberts}

\author{Timothy P. Roberts}
\affil{Department of Physics, Durham University, South Road, Durham, DH1 3LE, United Kingdom}
\email{t.p.roberts@durham.ac.uk}




\begin{abstract}
Ultraluminous X-ray sources (ULXs) are amongst the most intriguing of
X-ray source classes.  Their extreme luminosities - greater than
$10^{39} \rm ~erg~s^{-1}$ in the 0.3 -- 10 keV band alone - suggest
either the presence of black holes larger than those regularly
encountered in our own Galaxy (the Galactic centre excepted), or
sources apparently radiating well above the Eddington limit.  We
review the insights afforded us by studies of their X-ray emission,
focussing on what this reveals about the underlying compact object.
In particular, we discuss recent deep observations of ULXs by the {\it
XMM-Newton\/} observatory, and how the unprecedented data quality
provided by this mission is starting to discriminate between the
different physical models for these extraordinary X-ray emitters.
\end{abstract}


\keywords{black hole physics -- X-rays: binaries -- X-rays: galaxies
}

\section{Introduction}
\label{intro}

The first imaging observations of galaxies beyond our local group were
conducted by the {\it Einstein\/} observatory in the period 1978 -
1981, and revealed a somewhat unexpected result: many galaxies hosted
one or more extra-nuclear sources with X-ray luminosities well in
excess of those typically observed in our own galaxy and its nearest
neighbours \citep{Fab89}.  At the time it was rather presciently
acknowledged that if these were individual X-ray sources, their X-ray
luminosity was difficult to explain without invoking massive black
holes, or super-Eddington emission \citep{FabT87}.

Throughout the 1990s the {\it ROSAT\/} mission observed hundreds of
nearby galaxies, detecting many more of these sources
\citep{ColMush99,RW00,CP02,LiuBreg05}.  These observations revealed
little of the actual nature of these objects, except that a fraction
can be associated with recent supernovae, for example SN 1986J in NGC
891 \citep{BregPil92}.  However, subsequent analysis of samples of
these objects has given us some insights into their demographics, for
example with estimates of between only 1 in 8 and 1 in 4 major galaxies
hosting one of these X-ray bright objects with an observed luminosity
in excess of $10^{39} \rm~erg~s^{-1}$ \citep{PtakC04,LBI06}.  {\bf It
is this extreme luminosity threshold, in combination with an extra-nuclear location,
that we use to define this class of so-called ``ultraluminous X-ray
sources'' (ULXs)}.

A clearer insight into the nature of the majority of ULXs was provided
by {\it ASCA\/} observations in the years around 2000.  Most notably,
the novel wide bandpass CCD spectroscopy afforded by {\it ASCA\/}
allowed the spectra of many ULXs to be measured over the $0.5 - 10$
keV range for the first time.  This revealed some of them to be well-fitted by
the multi-colour disc blackbody model, that describes the
optically-thick thermal X-ray emission of an accretion disc around a
black hole \citep{Maki00}.  Additionally, multi-epoch observations of some 
ULXs saw them apparently transiting between spectral states described
by either the multi-colour disc blackbody model or a power-law,
similar to the transition between low- and high-states seen in
Galactic black hole X-ray binaries \citep{Kubota01}.  This gave the
first strong corroborating evidence - on top of their extraordinary
luminosities - that most ULXs are accreting black holes.  However,
these results were not without problems, notably that the disc
temperatures measured by \citet{Maki00} were far too hot to be
reconciled with the masses suggested by the high luminosities of the
sources (see below).  It was suggested that this could be explained by
rapidly spinning (Kerr metric) black holes, in which the inner edge of
the accretion discs are far closer to the black hole - and therefore
hotter - than in the non-rotating (Schwarzschild) case.

Our understanding of the nature of ULXs has increased immensely over
the last $\sim$ half decade due to the excellent capabilities of the
{\it Chandra\/} and {\it XMM-Newton} observatories, and follow-up
studies across the range of the electromagnetic spectrum.  In this
paper we mainly concentrate on contributions to the understanding of
ULXs garnered from spectroscopic and timing studies using the European
Photon Imaging Camera (EPIC) on {\it XMM-Newton}.  However, first we
will summarise some of the main arguments relating to the nature of
ULXs.

\subsection{A new class of black holes?}
\label{intro:a}

The Eddington limit for the maximum radiative luminosity possible from
the spherical accretion of matter\footnote{This equation is strictly
correct only for ionised hydrogen; the accretion of helium and/or
heavier elements will raise this limit.} on to a black hole can be
expressed as

\begin{equation}
\centering
L_{\rm Edd} = 1.3 \times 10^{38} (M/\msolar) \rm ~erg~s^{-1}
\label{leddeqn}
\end{equation}

where $M$ is the mass of the accreting object in solar masses
\citep{Maki00}.  Hence, for an object obeying the Eddington limit, at
a luminosity of $10^{39} \rm ~erg~s^{-1}$ its mass must be $\gtapprox
7.7 \msolar$.  For an accretion rate of $\sim 10$ per cent of that
required to reach the Eddington limit - a fairly typical accretion
rate for a high-state black hole - this means that a $\sim 77 \msolar$
black hole is required for the source to be emitting at $10^{39} \rm
~erg~s^{-1}$.  {\bf So, if ULXs obey the Eddington limit, they must
contain massive black holes.}  But how massive?

Dynamical friction arguments imply that these sources cannot be
misplaced super-massive black holes, sitting outside the nuclei of the
host galaxies, as such massive objects should sink to the centre of
the galaxies in a Hubble time \citep{TOS75}.  However, the Eddington
limit argument also rules out the stellar remnant black holes that we
know of in our own galaxy, with masses in the range $3\msolar < M_{\rm
BH} < 18\msolar$ \citep{McClRem06}, for all but the mildest of ULXs
($L_{\rm X} \ltapprox 2.3 \times 10^{39} \rm ~erg~s^{-1}$ at the
Eddington limit).  Indeed, \citet{FK01} calculate that the vast
majority of black holes formed from the evolution of a single massive
star will have mass $< 20\msolar$, clearly inadequate to power the
brighter end of the ULX population (if obeying the Eddington
limit).\footnote{\citet{FK01} do note that it is possible that very
massive, low metallicity stars leave a sufficiently massive remnant
core after the wind mass-loss phase ($\gtapprox 42\msolar$) to
collapse directly to a massive black hole.  However such objects would
be comparatively rare.}  These limits led to the suggestion that ULXs
may be the first observational evidence for a new, $\sim 10^2 - 10^5
\msolar$ {\it intermediate-mass} class of accreting black holes
(IMBHs) \citep{ColMush99} (see also \citet{MC04} for more on IMBHs).

The strongest supporting evidence in favour of IMBHs in ULXs comes
from the high signal-to-noise broad-band X-ray spectroscopy enabled by
{\it XMM-Newton\/}\footnote{Although similar results have been
obtained with {\it ASCA\/} \citep{ColMush99} and {\it Chandra\/}
\citep{Kaaret03,RC03,Roberts04}.}.  In particular, \citet{MFMF03}
showed that the spectra of two ULXs in NGC 1313 could be well fitted
by the same absorbed multi-colour disc blackbody plus power-law
continuum model that is commonly used as the empirical model to fit
Galactic black hole binaries, with the key difference being a lower
disc temperature in ULXs than Galactic black holes (0.1 -- 0.3 keV
versus $\sim 1$ keV, respectively).  This is crucial because, for a
fixed accretion rate, the temperature of the inner edge of a standard
accretion disc scales with the black hole mass as $T \propto
M^{-0.25}$, i.e. a cooler disc implies a bigger black hole.  In fact,
the black hole masses for the ULXs in NGC 1313 were estimated to be of
the order $\sim 1000 \msolar$.  Many ULX spectra were quickly shown to
agree with this result \citep{MFM04a,Cropper04,Dewangan04,Roberts05},
with \citet{MFM04b} demonstrating that such sources lie in a different
region of disc luminosity - disc temperature space than Galactic black
holes, emphasizing their potentially different natures.

Other factors also argue for the presence of IMBHs in at least some
ULXs.  For example, X-ray timing characteristics such as the detection
of quasi-periodic oscillations (QPOs) in the X-ray fluctuation Power
Spectral Densities (PSDs) of some ULXs argue that their emission is
isotropic, supportive of IMBHs assuming that the Eddington limit is
not exceeded (see Section~\ref{sec:2} for more details).  Similarly,
simple photon counting arguments for high-excitation optical line
emission regions near ULXs make the same argument
\citep{PM02,Kaaret04}.  The source for which most evidence stacks up
is M82 X-1, which through a combination of its extreme luminosity
($L_{\rm X, peak} \sim 10^{41} \rm ~erg~s^{-1}$), co-location with the
young, dense stellar cluster MGG 11, and QPO detections is the best
known candidate for an IMBH \citep{Kaaret01,SM03,PZ04,Mucci06}.
However, it is possible this source is an atypical ULX; it may be the
nucleus of an accreted dwarf galaxy \citep{KD05}.

\subsection{The problem(s) with IMBHs}
\label{intro:b}

Unfortunately, ULXs as a population are not trivially explained by the
presence of IMBHs.  There are in fact many arguments as to why the
majority of ULXs cannot be IMBHs, or at least IMBHs of the size
inferred from cool accretion discs ($\sim 1000 \msolar$).  Two
arguments stand out as the principle reasons the ULX population is not
dominated by (large) accreting IMBHs.  Firstly, the luminosity
function of X-ray sources in galaxies (XLF) has an unbroken power-law
form for 5 decades up to a luminosity of $\sim 2 \times 10^{40} \rm
~erg~s^{-1}$ \citep{GGS03,Swartz04}.  This break occurs at $\sim 10$
per cent of the Eddington luminosity for the $\sim 1000 \msolar$ black
holes inferred from ULX spectroscopy.  This is extremely troublesome
for a ULX population dominated by these large IMBHs, as they would not
only have to contrive to take over the XLF smoothly from Galactic
black holes at $\sim 10^{39} \rm ~erg~s^{-1}$, but then cease
accreting at 10 per cent of Eddington.  No other accreting source
class behaves in this manner.  This instead argues that ULXs are
dominated by black holes of mass up to $\sim 100 \msolar$ (or less if
the Eddington limit can be exceeded).

The second strong argument against IMBHs comes from the association of
ULXs with star formation.  Early observations with {\it Chandra\/}
revealed that starburst galaxies have populations of multiple ULXs
\citep{FZM01,Lira02,Roberts02}, an unusual result given the expectation
of less than one in four galaxies on average possessing even one ULX.
The obvious conclusion from this is that the ULXs are intrinsically
linked to the ongoing star formation occuring in those galaxies.
However, the direct co-location of ULXs with the star formation, most
notably seen in the Cartwheel galaxy \citep{Gao03}, implies that they
must be (relatively) short-lived, which requires successive
generations of ULXs to be formed over the duration of the star
formation event.  \citet{King04} pointed out that if these ULXs were
all large IMBHs, then an infeasibly large proportion of the available
star forming mass would end up in the form of IMBHs.  Hence the
majority of ULXs in star forming regions cannot be powered by IMBHs.

\subsection{The most extreme stellar-mass black holes?}
\label{intro:c}

If ULXs in starburst galaxies are not IMBHs, then what are they?  The
obvious solution is to turn to a class of objects we would expect to
find there anyway: high-mass X-ray binaries.  The problem then becomes
one of making such objects appear as ULXs.  Assuming Galactic black
hole masses for these objects (i.e. $M < 20 \msolar$) one then needs
to either make such objects {\it actually\/} break the Eddington
limit, or to make them {\it apparently\/} exceed it.  Lower-luminosity
and/or Eddington-limited objects could appear as bright ULXs due to
beaming, either through relativistic boosting of their X-ray emission
along our line-of-sight \citep{Kording02} or through collimation of
their radiation - probably by a geometrically-thick accretion disc -
such that it only escapes into a fraction of the sky
\citep{King01}.\footnote{QPO detections and high-excitation optical
line measurements argue against all but the mildest forms of beaming
in some ULXs.  Additionally, the lack of detectable radio emission
and/or rapid, high amplitude X-ray variability in most ULXs argues
against relativistic beaming - though see \citet{Krauss05} for a
possible counter-example.}  Alternatively, models have been suggested
whereby actual super-Eddington luminosities are achieved and
maintained, at factors $\ltapprox 10$ above the Eddington limit
\citep{Begelman02,Ebisawa03,HD07}.

Regardless of the processes involved, a very basic requirement of most
models is that sufficient fuel is available for the super-Eddington
mass transfer rates needed in ULXs.  \citet{Rappa05} show that this is
indeed the case for high-mass X-ray binary systems containing a
stellar-mass black hole and a massive donor star, that can in fact
sustain super-Eddington mass transfer over a very large fraction of
their lifetimes.  Other authors suggest that the possible
hyper-Eddington mass transfer rates in SS 433-like objects fuel ULXs
\citep{BKP06,Pout07}.  Furthermore, where optical stellar counterparts
to ULXs have been identified, primarily by {\it HST\/}, they tend to
be blue and of an appropriate magnitude for the young, massive stars
required to fuel ULXs \citep{Roberts01,Liu04,Kuntz05}.  This provides
compelling support to the argument that ULXs are high-mass X-ray
binaries.

Finally, it is reassuring to know that the Eddington limit is broken
in practise in Galactic black holes - \citet{McClRem06} give several
examples, most notably that of GRS 1915+105.  This source is
sufficiently luminous (at $\sim 10^{39} \rm ~erg~s^{-1}$) to appear as
a ULX if viewed from outside our galaxy, and (for its known black hole
mass) has consistently displayed peak luminosities in excess of the
Eddington limit over the $\sim 15$ years of its outburst to date
\citep{Done04}.  As the Eddington limit is exceeded in known sources,
there is no reason this cannot also be occuring in ULXs.

We are therefore left with the situation where most ULXs could be
explained by stellar-mass black holes that are either super-Eddington,
or subject to some sort of beaming.  However, it is still difficult to
reconcile the most extreme ULXs - those above $10^{40} \rm
~erg~s^{-1}$ - with simple stellar-mass black hole systems.  Larger
black holes would still provide an obvious solution.  But are they
really the $\sim 1000 \msolar$ black holes inferred from ULX
spectroscopy?

\section{X-ray spectroscopy}
\label{sec:1}

\subsection{The IMBH model, and other solutions}
\label{sec:1a}

As we have already discussed, the strongest support for ULXs
containing IMBHs comes from X-ray spectroscopy, and in particular the
good fits obtained to ULX spectra using the same multi-colour disc
plus power-law continuum model used for Galactic black holes.  To date
more than 10 ULXs with decent {\it XMM-Newton\/} spectra have been
shown to have their spectral fits improved substantially (compared to,
say, a simple absorbed power-law model) by the use of this model, in
many cases producing a statistically acceptable fit to the data.  An
ubiquitous feature of these fits is a cool disc which, as we have
already seen, when interpreted at face value implies the presence of
an IMBH with mass of the order $\sim 1000 \msolar$.

\begin{figure}[t!]
\begin{center}
\includegraphics[width=0.33\textwidth, angle=270]{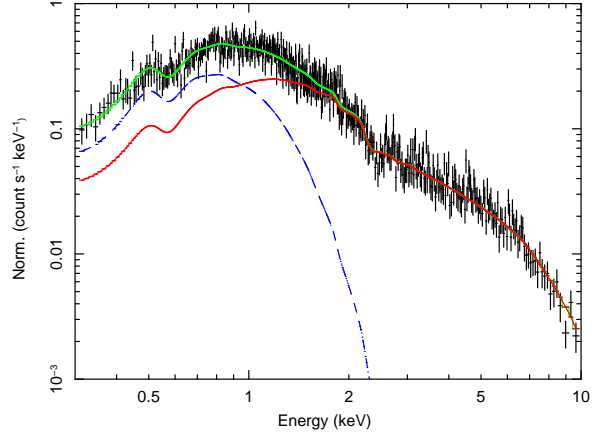}
\end{center}
\caption{{\it XMM-Newton\/} EPIC-pn spectrum of NGC 1313 X-1,
obtained on 2000 October 17, and re-processed with \textsc{sas}
version 6.5.0.  The data points are shown in black, with the
best-fitting IMBH model in green.  The contributions to this model of
the multi-colour disc black body component (with $kT_{\rm in}
\sim 0.2$ keV) and the power-law continuum ($\Gamma \sim 1.7$) are
shown by blue (dashed) and red lines respectively.  The X-ray emission
detected by {\it XMM-Newton\/} is clearly dominated by the power-law
component, and not the disc.}
\label{imbhmod}
\end{figure}

But should we take this mass at face value?  There is good reason not
to.  The mass estimates deriving from the multi-colour disc blackbody
model assume that this is the dominant emission component in the X-ray
spectrum (essentially, that the black hole is in the ``high'' state).
Unfortunately, in ULXs fit by this ``IMBH model'' (which here
specifically refers to an empirical model composed of a cool disc plus
harder power-law) it is very evidently not.  We demonstrate this
graphically for the IMBH candidate ($L_{\rm X} > 10^{40} \rm
~erg~s^{-1}$) NGC 1313 X-1 in Fig~\ref{imbhmod}, where it is obvious
that the power-law dominates the flux within the {\it XMM-Newton\/}
bandpass.  In fact, the disc typically emits no more than 20 per cent
of the 0.3 -- 10 keV flux of ULXs in this model \citep{SRW06}.  Worse
still, the power-law slopes measured by this model are somewhat on the
low side for the classic high state ($\Gamma \sim 1.6 - 2.5$ for IMBH
models, compared to $\Gamma \sim 2.1 - 4.8$ in Galactic high state
sources; cf. \citet{McClRem06}).  This makes black hole masses
obtained by this method (at the very least) questionable.

Furthermore, the IMBH model is not the only model that fits ULX
spectra.  Several sources have been identified in which a variant of
this model, where the disc component fits to the hard end of the
spectrum, provides a far superior fit \citep{SRW04,Foschini04,FK05}.
As discussed by \citet{Roberts05} this variant of the empirical black
hole spectrum model does not provide a physical model for the X-ray
emission - for example there cannot be sufficient photons present in
the vicinity of the black hole to produce the dominant soft power-law
through Compton up-scattering.  However, the crucial point is that
there is distinct curvature (which can also be described as a spectral
``break'') present above 2 keV in these data, for which there must be
a physical explanation.

\subsection{Re-evaluating ULX spectra from XMM-Newton}
\label{sec:1b}

Given the question marks about the IMBH model, and this second,
``inverted'' model that fit to some ULXs, we set out to examine the
best available {\it XMM-Newton\/} ULX datasets (from the beginning of
2005).  Specifically, in \citet{SRW06} we set out to ask the questions

\begin{itemize}
\item
How easy is it to distinguish the IMBH and inverted models given the
available quality of data?
\item
With what frequency do these models work for the available data?
\item
Can we say anything about the physics of the inverted model?
\end{itemize}

\begin{figure}[t!]
\begin{center}
\includegraphics[width=0.33\textwidth, angle=270]{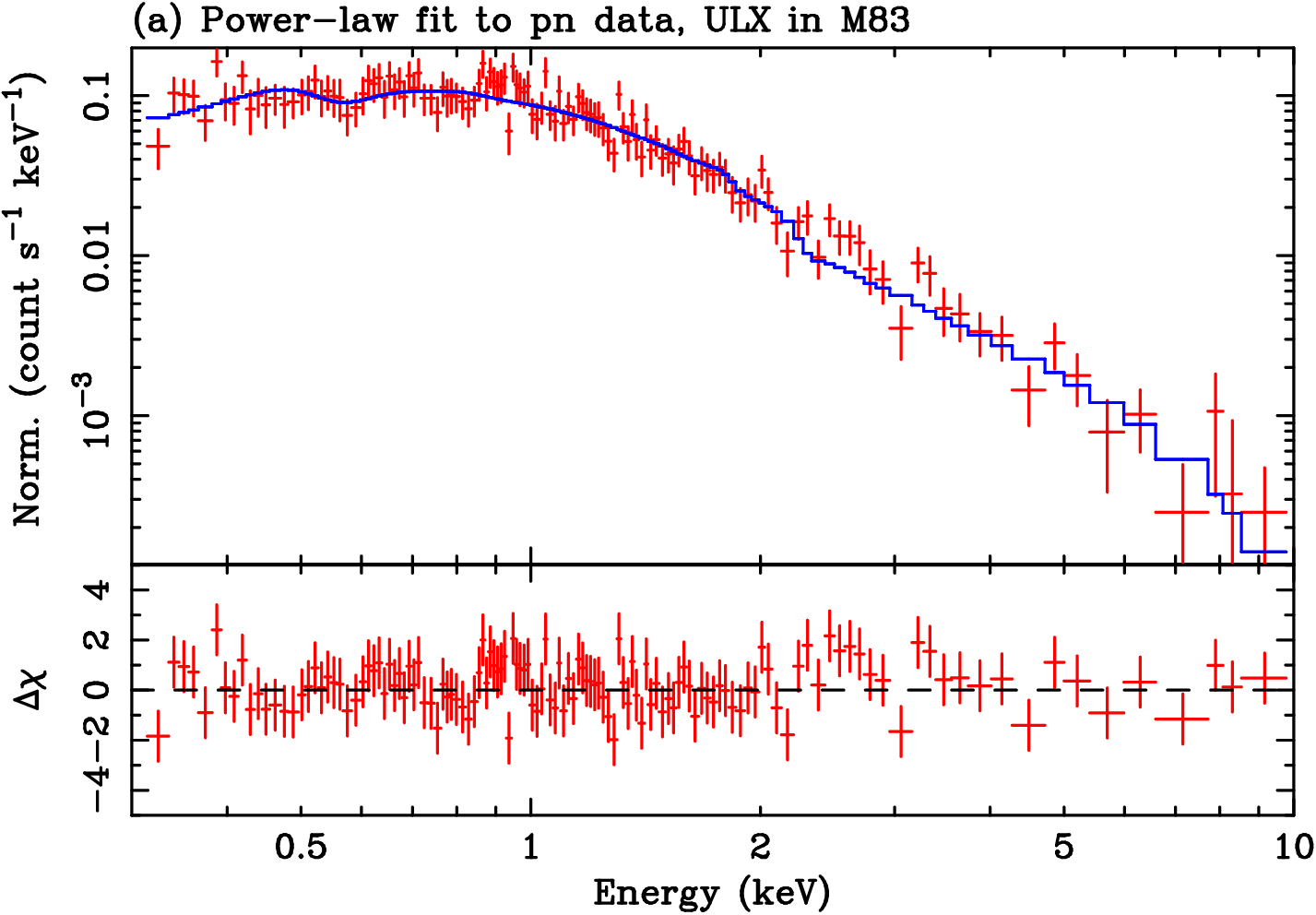}\vspace*{2mm}
\includegraphics[width=0.33\textwidth, angle=270]{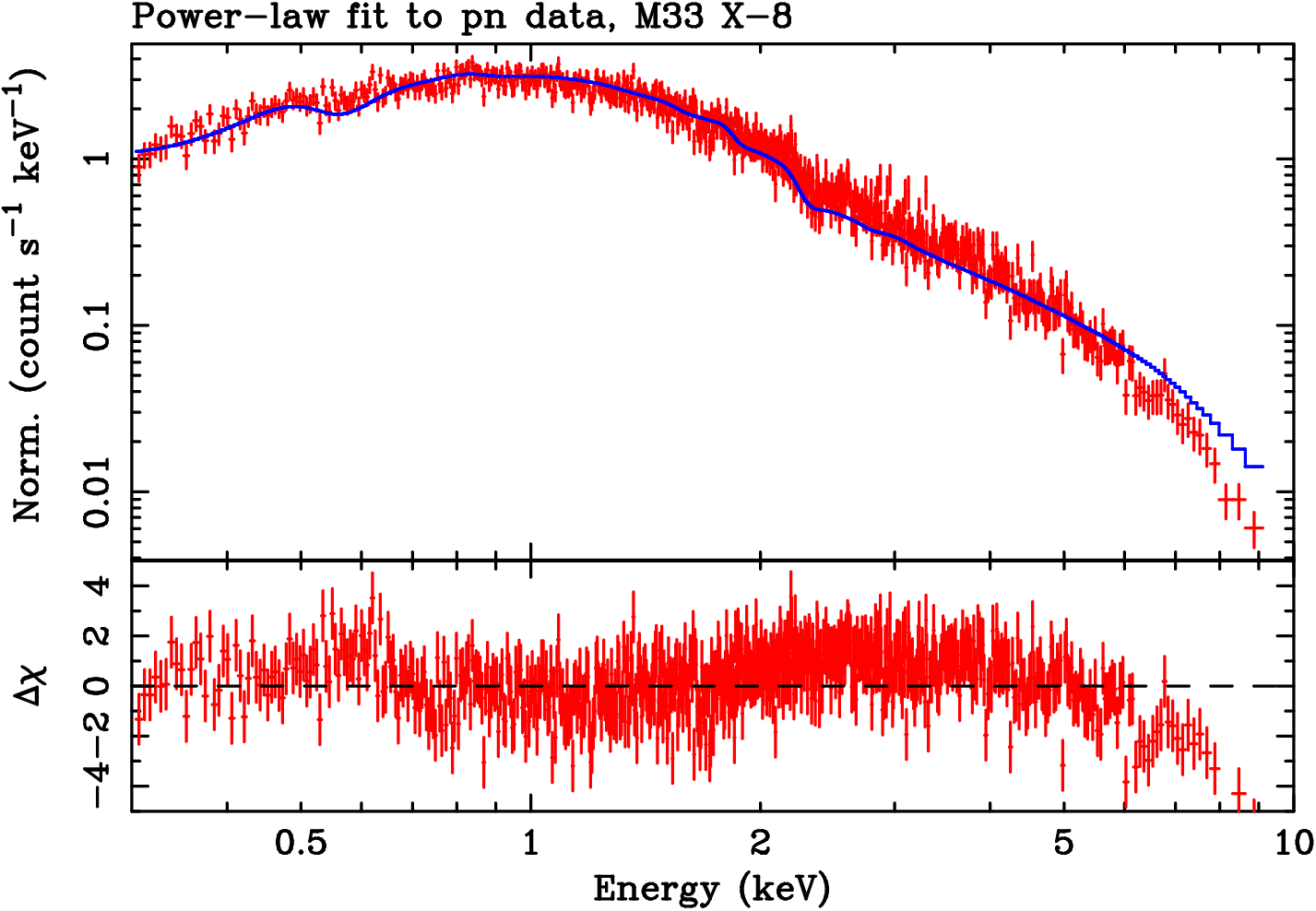}
\end{center}
\caption{{\it XMM-Newton\/} EPIC-pn data for two ULXs in the
\citet{SRW06} sample.  In both cases we show the data (in red) and
best-fitting power-law continuum model (in blue) in the top panel,
with the $\Delta\chi$ residuals for the fit shown in the bottom.  The
ULX in M83 was adequately fit by a power-law continuum, whereas M33
X-8 was not.}
\label{specqual}
\end{figure}

To do this we selected data sets with at least a few thousand counts
(EPIC-pn and MOS combined) per ULX.  We show examples of the low and
high end of the data quality for ULXs in our sample in
Fig~\ref{specqual}.  The data was of sufficient quality to
statistically rule out simple multi-colour disc blackbody fits to all
the sources, and power-law continua in 8/13 cases.

As simple models were inadequate for all but the poorest data, we next
attempted empirical two-component models, beginning with the IMBH
model.  This provided acceptable fits to 8/13 sources, which all had
the classic $\sim 0.1 - 0.3$ keV cool disc signature.  Unfortunately,
they all also displayed the problems inherent in mass measurements
from this model, i.e. dominant, hard power-law continua.  We then
attempted fits using the inverted model.  This also provided good fits
to 8/13 data sets, parameterised by $\Gamma \sim 2.5 - 4.3$ power-law
photon indices, $kT_{\rm in} \sim 0.9 - 2.7$ keV inner-disc
temperatures, and a very roughly 50/50 split between the flux
contribution of the two components in the {\it XMM-Newton\/} band.
Interesting, the six lowest quality data sets provided acceptable fits
to both models, demonstrating that either very high quality X-ray
data, or a secondary diagnostic, is required to distinguish the
models.

This key diagnostic is found at energies above 2 keV, where
disc-domination leads to distinct curvature in the spectrum, whereas a
dominant power-law has an unbroken spectrum.  We tested for this
characteristic signature by comparing power-law fits to broken
power-law fits on the $> 2$ keV data for each ULX.  In total we found
8 ULXs showing evidence for breaks, at significance levels between $3
- 10 \sigma$.  Three of these were expected, as they were from sources
clearly better fit by the inverted model.  However, five sources that
were either ambiguous or well-fitted by the IMBH model also showed a
significant break.  Of the remaining (unbroken) source fits, in three
cases this may be attributable to very poor data quality above 2 keV.
This simple test therefore demonstrates that spectral breaks are
present in the majority of ULXs (notably, across the whole range of
luminosity) where the data quality above 2 keV is sufficient to detect
them.\footnote{A by-product of this spectral break/curvature is that
the best empirical fits are provided by models composed of two thermal
components, in particular a combination of a blackbody with a
multi-colour disc blackbody model - see \citet{SRW06} for more
details.}

In order to investigate the physics of this break further, we
attempted spectral fits using a physically self-consistent accretion
disc plus Comptonised corona model, specifically using an absorbed
DISKPN + EQPAIR model in \textsc{xspec}.  Where appropriate,
parameters were tied to assumed values (based on experience with
Galactic systems), so that the model had only one more degree of
freedom than the empirical (two-component) models, but sufficient
scope for variation within the model parameters was allowed such that
the outcomes were not pre-judged.  This model gave superior fits to
the empirical models, with 11/13 ULXs providing
statistically-acceptable fits (with one more only marginally
unacceptable).  Two remarkable characteristics were common to most
fits: firstly, the ULXs still showed apparently cool discs; but
secondly, the coronae (in 9/12 cases) appeared {\bf optically-thick}
(with $\tau > 8$ - in several cases much higher).  It is this optical
thickness in the corona that is responsible for the curvature in the 2
- 10 keV spectrum.  Furthermore, this distinguishes ULXs from galactic
black holes, that do not {\it typically\/} show such coronae.  The
reason their coronae are empirically modelled by power-laws is that
they are optically-thin.  This therefore suggests ULXs are operating
in a different accretion mode to the classic states of Galactic black
holes.

\subsection{Possible scenarios}
\label{sec:1c}

So what is happening in these sources?  One physical scenario that
could lead to the production of both a cool disc component and an
optically-thick corona is suggested by \citet{Zhang00}, based on an
analogy with the solar corona.  This ``sandwich'' model consists of a
cool inner accretion disc, seeding an outer, warm ($\sim 1$ keV) and
optically-thick ($\tau \sim 10$) accretion disc layer with $\sim 0.2$
keV photons.  This could readily provide both components we derive
from our spectra, though the geometry required to see the cool photons
through the warm layer is problematic.  A second scenario is based on
observations of a strongly-Comptonised very high state observed in the
Galactic black hole XTE J1550-564 by \citet{Done06}.  In this source
they also detect a cooler-than-expected disc, alongside an optically
thick corona (though neither phenomena are as extreme as detected in
the ULXs), and suggest that this is due to energetic coupling of the
inner-disc with the corona.  In such a system, the energy released by
the extreme accretion rate is sufficient to launch an optically-thick
corona, which obscures (and, through extracting the launch energy,
cools) the central regions of the disc.  One therefore predominantly
sees the cooler outer regions of the accretion disc, in addition to
the optically-thick corona.  As this state occurs at a very high
accretion rate in XTE J1550-564, this suggests that ULXs operate at
similar - or higher - accretion rates.  This must be at around the
Eddington limit, suggestive of black hole masses up to $\sim 100
\msolar$.

Further work has now revealed this $> 2$ keV break in other ULXs - for
example in M82 X-1 \citep{Okajima06} and Ho IX X-1 (aka M81 X-9)
\citep{Dewangan06}.  Other ideas have been also been postulated for
its origin, and the full observed X-ray spectrum for ULXs.  Several
authors have discussed ULXs in the context of slim disc models
\citep{Watarai01}.  In particular, they note that as the Eddington
limit is approached, the structure of the accretion disc should
change.  This would manifest itself as a change in the model disc
profile, $T(r) \propto r^{-p}$, where standard discs have $p = 0.75$,
and slim discs $p = 0.5$.  Recent work where a variable disc profile
model is fit to ULX spectra does indeed show values of $p \sim 0.6$,
suggestive of slim discs \citep{Vier06,mizuno07}.  A second idea, put
forward by \citet{Gonc06}, draws from observations of AGNs with
outflows.  In their model ULXs have an intrinsic very high state
spectrum (steep power-law form, i.e. $\Gamma > 2.5$), that is modified
by absorption from material in an ionised fast outflow.  This
effectively takes a ``bite'' out of the spectrum in the $\sim 1 - 4$
keV range, resulting in the apparent soft excess and $> 2$ keV
spectral break features.  We note that the post-break power-law slopes
measured by \citet{SRW06} are indeed consistent with a very high state
spectrum.  Finally, work by \citet{Pout07} has related ULX spectra to
the expected spectra from super-critically accreting black holes
viewed with the accretion disc close to face-on.  In such sources - SS
433 may be an example, viewed edge-on - the break originates from a
direct view of the spectrum of the inner-regions of a hot accretion
disc.


All these models have a common theme - accretion at around or above
the Eddington rate.  This strongly suggests black holes of $\ltapprox
100 \msolar$, rather than the $\sim 1000 \msolar$ IMBHs formerly
proposed.

\subsection{Spectral variability}
\label{sec:1d}

Snapshot observations, though useful, can only tell us so much about
accreting sources.  As 10 years of {\it RXTE\/} observations have
shown, for example with Galactic black holes, further progress can be
made by considering how these observed spectra change with time
\citep{McClRem06}.  As ULXs are in general too X-ray faint for
monitoring missions like {\it RXTE\/} (with the exception of M82 X-1 -
see \citet{KSL06}), the best one can do is to use monitoring campaigns
of snapshot observations separated by days -- months.

\begin{figure}[t!]
\begin{center}
\includegraphics[width=0.35\textwidth, angle=270]{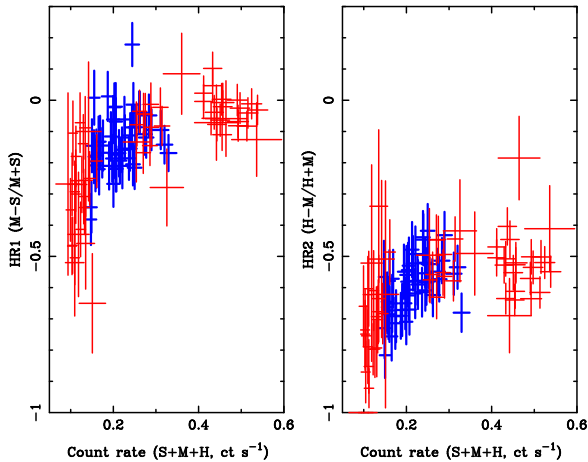}
\end{center}
\caption{{\it Chandra\/} ACIS-S hardness ratio - count rate plots for
NGC 5204 X-1.  The bands used are $S = 0.3 - 1$ keV; $M = 1 - 2$ keV;
$H = 2 - 8$ keV.  Blue data points are from an initial 50-ks
observation, with data from ten 5-ks follow-ups shown in red.  The ULX
spectrum clearly hardens as its flux increases.}
\label{5204HRs}
\end{figure}

Unfortunately, to date very few campaigns have been pursued.  Where
they have been undertaken, sources are observed to behave in two
distinct ways.  Some ULXs behave as would be expected from classic
Galactic black hole behaviour, i.e. they get spectrally softer as
their flux increases.  In contrast, other ULXs behave in the opposite
fashion - they get harder
\citep{Fab03,Dewangan04,Jenkins04,FK06a,FK06b,Soria07a}.  We have
undertaken such a campaign for the ULX NGC 5204 X-1 using {\it
Chandra\/} data, with observations separated by days -- weeks, and
find that it behaves in the latter mode \citep{Roberts06} - see
Fig~\ref{5204HRs}.  In particular, we show that this behaviour can be
modelled as changes in the temperature of the optically-thick
component of a cool accretion disc plus thick corona model, with the
corona heating up as the luminosity of the ULX increases.  This
behaviour was also seen in the strongly-Comptonised very high state of
XTE J1550-564 \citep{KD04,Soria07b}, confirming the viability of the
optically-thick corona model for this ULX.

%
%


\section{X-ray timing}
\label{sec:2}

Perhaps the most fundamental timing signal from a ULX is an X-ray
periodicity, as it can be used as a first step towards establishing
the orbital characteristics of the underlying binary system.  However,
such measurements are not very common, with the best detections coming
in a couple of cases where eclipses have been found
\citep{Bauer01,David05,Fab06}.  Though other claims have been made,
they suffer from a lack of data - most are based on $\ll 10$ cycles,
and require confirmation through further observations.  In the absence
of periodicities, the most useful timing diagnostic is Power Spectral
Density (PSD) measurements for ULXs.

\subsection{Power Spectral Densities for ULXs}
\label{sec:2a}

\begin{figure}[t!]
\begin{center}
\includegraphics[width=0.35\textwidth, angle=270]{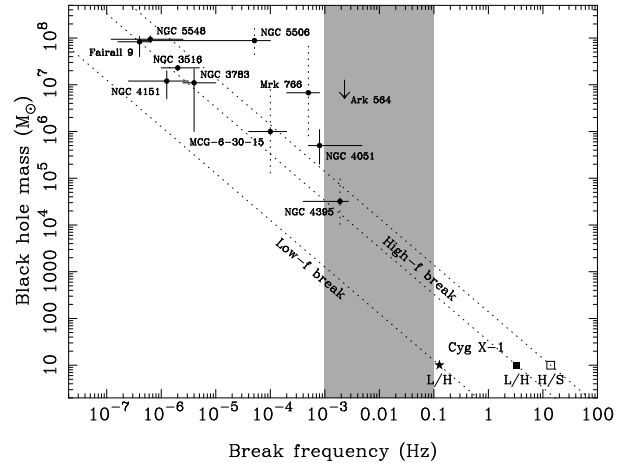}
\end{center}
\caption{A simplified linear relationship between black hole mass and
PSD break frequency for Cygnus X-1 and a number of AGN.  This plot is
a reasonable approximation for the $T_{\rm break} \propto M_{\rm
BH}^{1.12} \dot{m}_{\rm Edd}^{-0.98}$ relationship found by
\citet{McHardy06}, assuming sources at the Eddington limit (where
$T_{\rm break}$ is the break timescale, $M_{\rm BH}$ is the black hole
mass, and $\dot{m}_{Edd}$ is the accretion rate in Eddington units).
Note that sub-Eddington accretion rates move the break to the left on
this plot (and vice versa). We shade the region for which {\it
XMM-Newton\/} observations are sensitive, showing that this coincides
with IMBH-like masses. {\it Courtesy S. Vaughan.\/}}
\label{PSDbreaks}
\end{figure}

The shape of the fluctuation PSD for ULXs is potentially a very
powerful tool for ULXs.  In particular, the characteristic frequency
of breaks in the PSD slope can be used to infer masses based on a
direct scaling of properties between Galactic black holes and AGN -
see \citet{McHardy06}.  In Fig~\ref{PSDbreaks} we demonstrate this
concept using a simple linear scaling of black hole mass to break frequency,
and show that the 10 - 1000 s timescales over which {\it XMM-Newton\/}
observations are most sensitive to measuring breaks matches well with
that in which one might find IMBHs.  

Unfortunately, ULXs do not generally show much short-term variability
\citep{Swartz04}, with very few examples displaying sufficient variability
power to establish a PSD from \citep{FK05}.  However, some
measurements have been made.  \citet{Cropper04} detected a putative
break at 28 mHz in the PSD of NGC 4559 X7, which they suggest supports
the case that it harbours a $\sim 1000 \msolar$ black hole, although
the presence of this spectral break is now disputed \citep{Barnard07}.
Another detection of a break frequency was made by \citet{Soria04},
who found a break at 2.5 mHz in the PSD of NGC 5408 X-1.  When the
shape of the PSD either side of the break was considered, the authors
derived a mass of $\sim 100 \msolar$ for this ULX.  Finally,
\citet{DTG06} find a break at $\sim 34$ mHz in the PSD of M82 X-1,
from which they infer a mass in the $25 - 520 \msolar$ range for the
underlying black hole.

In 2004 we were awarded a 100-ks {\it XMM-Newton\/} observation in
order to derive a PSD for the nearby, luminous ULX Holmberg II X-1
\citep{Goad06}.  However, we found that it displayed no strong
variability during the observation.  In fact, through a PSD analysis
we were able to demonstrate that the fractional variability amounted
to less than a few per cent {\it rms\/} over time scales of minutes to
hours. This variability power is less than that observed in classic
high and very high state (red noise) PSDs.  The PSD could instead be
consistent with the low/hard state, but the energy spectrum for Ho II
X-1 is closer to a very high state spectrum (in fact, it displays a
cool disc plus optically-thick corona spectrum).  One solution is that
the source could be in a state similar to the '$\chi$'-class of
GRS1915+105, in which the source is in a very high state, but its PSD
is band-limited (i.e. all its variability power is limited to a narrow
frequency band).  As we do not see this band-limited variability, it
is likely to be at higher frequencies than we are sensitive to in our
PSD.  This means the black hole must be small.  In fact, we calculate
a limit of $\ltapprox 100 \msolar$ for Ho II X-1 from the lack of
variability.

\subsection{QPOs}
\label{sec:2b}

A second feature of PSDs that has diagnostic potential for ULXs are
quasi-periodic oscillations (QPOs).  In fact, the detection of 3:2
ratio twin-peak high-frequency QPOs at $\sim 1$ Hz would be strong
evidence for the presence of an IMBH \citep{Abr04} (though scaling
from Galactic systems implies such a measurement is unlikely even in
the medium term).  Low frequency QPOs, on the other hand, have now
been detected in a handful of ULXs
\citep{SM03,Dewangan06,Mucci06,Stroh07}.  However, this type of QPO
does not provide a clear, unambiguous mass estimate.  For example, the
detection of a QPO in NGC 5408 X-1, even with the additional
diagnostics of a break frequency and a second (4:3 ratio) possible
QPO, could imply mass estimates anywhere in the range $100 - 1000
\msolar$ dependent upon the assumptions made \citep{Stroh07}.  Despite
this, QPOs are telling us one thing - as they are a coherent signal,
their detection rules out all but the mildest forms of beaming in ULXs
where they are present.

\section{Concluding remarks}
\label{sec:3}

New observational evidence is now pointing away from the
interpretation of ULXs as the $\sim 1000 \msolar$ black holes inferred
from simple cool disc plus power-law spectral models.  Putting to one
side the inherent problems with the dominant, hard power-law component
derived from this model, the crucial evidence in this matter is the
spectral break above 2 keV detected by \citet{SRW06}.  This is
completely unexpected and inexplicable in the context of the simple
$\sim 1000 \msolar$ black hole interpretation for ULXs.  Indeed, it
suggests physical characteristics that have more in common with
Galactic sources accreting at around the Eddington limit.  This
implies much smaller black hole masses - of the order $\sim 100
\msolar$ or less - for ULXs.

Are ULXs then stellar-mass black holes that are radiating at
super-Eddington rates, i.e. factors of $\ltapprox 10$ above the
Eddington limit for most ULXs, assuming masses of $< 20 \msolar$?
Though this is feasible, it is perhaps not necessary.  Almost all ULXs
could trivially be explained (at least in luminosity terms) by
accreting black holes with masses of a few tens of $\msolar$,
consistent with their spectra suggesting they are accreting at around
(including slightly above) the Eddington limit.  Interestingly,
results based on optical observations of ULX counterparts are
suggesting a similar conclusion.  For example, a possible radial
velocity variation in a He II line detected from the optical spectrum
of NGC 1313 X-2 by \citet{PGM06} suggests a small black hole mass.
Furthermore, irradiation models of donor stars in ULXs, when combined
with optical colours, suggest black hole masses of the order
$\ltapprox 100 \msolar$ \citep{Copp07}.  A final piece of the puzzle
is that the creation of such black holes may be possible in the young
stellar populations that we generally find ULXs co-habiting with, from
the merging of a binary composed of very massive early-type stars
\citep{Belcz06} (or, alternatively, see footnote 2).  This could yield
the black holes with masses of up to $\sim 100 \msolar$ that we are
potentially finding in ULXs.

Hence it now appears that we can tentatively conclude that the vast
majority of ULXs could harbour the slightly bigger cousins of the
Galactic black holes we are familiar with, rather than their far
larger distant relatives.  Should we call them IMBHs?  The definition
of this term is somewhat indistinct, with some authors quoting $20
\msolar$ as the lower limit for this class, and others starting at
$\sim 100 \msolar$.  The few times larger than stellar-mass black
holes that could be powering ULXs are precisely in this grey
area.  Perhaps a new artificial distinction between ``small''
($\ltapprox 100 \msolar$) and ``large'' IMBHs is required for
clarity's sake.

However, the mass of an underlying black hole is yet to be
conclusively determined for any individual ULX - this requires the
measurement of a dynamical mass function, which is non-trivial for
extra-galactic sources.  Furthermore, a small and very rare sub-group
of the most luminous ULXs - the ``hyperluminous X-ray sources'' with
$L_{\rm X} > 10^{41} \rm ~erg~s^{-1}$ \citep{Gao03,Wolter06,Mini06} -
defy easy explanation by anything other than large IMBHs, assuming
they are indeed accreting black holes in the host galaxies, and not
luminous supernovae or foreground/background objects.  The issue of
whether some ULXs could still constitute evidence for accretion onto
large IMBHs is therefore far from finished with.

\begin{acknowledgements}
TPR would like to thank his collaborators on the work presented here
-- most notably Ann-Marie Stobbart and Mike Goad -- and a select yet
too-numerous-to-mention group of workers in this and related fields
for many very useful conversations.  He would also like to apologise
for omitting to mention many other good pieces of work on ULXs due to
simple space and subject limitations.  Many of the results quoted in
this work are based on observations obtained with {\it XMM-Newton\/},
an ESA science mission with instruments and contributions directly
funded by ESA Member States and NASA.  Finally, thanks to Martin Ward
for presenting this work at the 5th Stromlo Symposium in TPR's
absence.
\end{acknowledgements}

\bibliographystyle{Spr-mp-nameyear}





\begin{thebibliography}{}

\bibitem[\protect\citeauthoryear{Abramowicz et al.}{2004}]{Abr04} 
Abramowicz M.A., Klu{\'z}niak W., McClintock J.E., Remillard R.A.,
2004, ApJ, 609, L63
\bibitem[\protect\citeauthoryear{Barnard et al.}{2007}]{Barnard07}
Barnard R., Trudolyubov S., Kolb U.C., Haswell C.A., Osborne J.P.,
Priedhorsky W.C., 2007, {\tt astro-ph/0703120}
\bibitem[\protect\citeauthoryear{Bauer et al.}{2001}]{Bauer01} 
Bauer F.E., Brandt W.N., Sambruna R.M., Chartas G., Garmire G., Kaspi
S., Netzer H., 2001, AJ, 122, 182
\bibitem[\protect\citeauthoryear{Begelman}{2002}]{Begelman02} 
Begelman M.C., 2002, ApJ, 568, L97
\bibitem[\protect\citeauthoryear{Begelman, King \& Pringle}{2006}]{BKP06} 
Begelman M.C., King A.R., Pringle J.E., 2006, MNRAS, 370, 399
\bibitem[\protect\citeauthoryear{Belczynski et al.}{2006}]{Belcz06} 
Belczynski K., Sadowski A., Rasio F.A., Bulik T., 2006, ApJ, 650, 303
\bibitem[\protect\citeauthoryear{Bregman \& Pildis}{1992}]{BregPil92} 
Bregman J.N., Pildis R., 1992, ApJ, 398, L107
\bibitem[\protect\citeauthoryear{Colbert \& Mushotzky}{1999}]{ColMush99}
Colbert E.J.M., Mushotzky R.F., 1999, ApJ, 519, 89
\bibitem[\protect\citeauthoryear{Colbert \& Ptak}{2002}]{CP02} 
Colbert E.J.M., Ptak A.F., 2002, ApJS, 143, 25
\bibitem[\protect\citeauthoryear{Copperwheat et al.}{2007}]{Copp07}
Copperwheat C., Cropper M., Soria R., Wu K., 2007, MNRAS, 376, 1407
\bibitem[\protect\citeauthoryear{Cropper et al.}{2004}]{Cropper04} 
Cropper M., Soria R., Mushotzky R.F., Wu K., Markwardt C.B., Pakull
M., 2004, MNRAS, 349, 39
\bibitem[\protect\citeauthoryear{David et al}{2005}]{David05} 
David L.P., Jones C., Forman W., Murray S.S., 2005, ApJ, 635, 1053
\bibitem[\protect\citeauthoryear{Dewangan et al.}{2004}]{Dewangan04} 
Dewangan G.C., Miyaji T., Griffiths R.E., Lehmann I., 2004, ApJ, 608, L57
\bibitem[\protect\citeauthoryear{Dewangan, Griffiths \& Rao}{2006}]{Dewangan06} 
Dewangan G.C., Griffiths R.E., Rao A.R., 2006, ApJ, 641, L125
\bibitem[\protect\citeauthoryear{Dewangan, Titarchuk \& Griffiths}{2006}]{DTG06} 
Dewangan G.C., Titarchuk L., Griffiths R.E., ApJ, 637, L21
\bibitem[\protect\citeauthoryear{Done \& Kubota}{2006}]{Done06} 
Done C., Kubota A., 2006, MNRAS, 371, 1216
\bibitem[\protect\citeauthoryear{Done, Wardzi{\'n}ski \& Gierli{\'n}ski}{2004}]{Done04} 
Done C., Wardzi{\'n}ski G., Gierli{\'n}ski M., 2004, MNRAS, 349, 393
\bibitem[\protect\citeauthoryear{Ebisawa et al.}{2003}]{Ebisawa03} 
Ebisawa K., Zycki P., Kubota A., Mizuno T., Wataria K., 2003, ApJ, 597, 780
\bibitem[\protect\citeauthoryear{Fabbiano}{1989}]{Fab89}
Fabbiano G., 1989, ARA\&A, 87, 27
\bibitem[\protect\citeauthoryear{Fabbiano et al.}{2006}]{Fab06}
Fabbiano G., et al., 2006, ApJ, 650, 879
\bibitem[\protect\citeauthoryear{Fabbiano \& Trinchieri}{1987}]{FabT87}
Fabbiano G., Trinchieri G., 1987, ApJ, 315, 46
\bibitem[\protect\citeauthoryear{Fabbiano et al.}{2003}]{Fab03} 
Fabbiano G., Zezas A., King A.R., Ponman T.J., Rots A., Schweizer F.,
2003, ApJ, 584, L5
\bibitem[\protect\citeauthoryear{Fabbiano, Zezas \& Murray}{2001}]{FZM01} 
Fabbiano G., Zezas A., Murray S.S., 2001, ApJ, 554, 1035
\bibitem[\protect\citeauthoryear{Feng \& Kaaret}{2005}]{FK05} 
Feng H., Kaaret P., 2005, ApJ, 633, 1052
\bibitem[\protect\citeauthoryear{Feng \& Kaaret}{2006a}]{FK06a} 
Feng H., Kaaret P., 2006, ApJ, 650, L75
\bibitem[\protect\citeauthoryear{Feng \& Kaaret}{2006b}]{FK06b} 
Feng H., Kaaret P., 2006, ApJ, 653, 536
\bibitem[\protect\citeauthoryear{Foschini et al.}{2004}]{Foschini04} 
Foschini L., Rodriguez J., Fuchs Y., Ho L.C., Dadina M., Di Cocco G.,
Courvoisier T.J.-L., Malaguti G., 2004, A\&A, 416, 529
\bibitem[\protect\citeauthoryear{Fryer \& Kalogera}{2001}]{FK01}
Fryer C.L., Kalogera V., 2001, ApJ, 554, 548
\bibitem[\protect\citeauthoryear{Gao et al.}{2003}]{Gao03} 
Gao Y., Wang Q.D., Appleton P.N., Lucas R.A., 2003, ApJ, 596, L171
\bibitem[\protect\citeauthoryear{Goad et al.}{2006}]{Goad06}
Goad M.R., Roberts T.P., Reeves J.N., Uttley P., 2006, MNRAS, 365, 191
\bibitem[\protect\citeauthoryear{Goncalves \& Soria}{2006}]{Gonc06} 
Goncalves A., Soria R., 2006, MNRAS, 371, 673
\bibitem[\protect\citeauthoryear{Grimm, Gilfanov \& Sunyaev}{2003}]{GGS03} 
Grimm H.-J., Gilfanov M., Sunyaev R., 2003, MNRAS, 339, 793
\bibitem[\protect\citeauthoryear{Heinzeller \& Duschl}{2007}]{HD07} 
Heinzeller D., Duschl W.J., 2007, MNRAS, 374, 1146
\bibitem[\protect\citeauthoryear{Jenkins et al.}{2004}]{Jenkins04} 
Jenkins L.P., Roberts T.P., Warwick R.S., Kilgard R.E., Ward M.J.,
2004, MNRAS, 349, 404
\bibitem[\protect\citeauthoryear{Kaaret et al.}{2003}]{Kaaret03} 
Kaaret P., Corbel S., Prestwich A.H., Zezas A., 2003, Science, 299, 365
\bibitem[\protect\citeauthoryear{Kaaret et al.}{2001}]{Kaaret01} 
Kaaret P., Prestwich A., Zezas A., Murray S., Kim D.-W., Kilgard
R., Schlegel E., Ward M., 2001, MNRAS, 321, L29
\bibitem[\protect\citeauthoryear{Kaaret, Simet \& Lang}{2006}]{KSL06}
Kaaret P., Simet M.G., Lang C.C., 2006, ApJ, 646, 174
\bibitem[\protect\citeauthoryear{Kaaret, Ward, \& Zezas}{Kaaret et
al.}{2004}]{Kaaret04} Kaaret P., Ward M.~J., Zezas A., 2004, MNRAS,
351, L83
\bibitem[\protect\citeauthoryear{King}{2004}]{King04} 
King A., 2004, MNRAS, 347, L18
\bibitem[\protect\citeauthoryear{King et al.}{2001}]{King01} 
King A., Davies M.B., Ward M.J., Fabbiano G., Elvis M., 2001, ApJ,
552, L109
\bibitem[\protect\citeauthoryear{King \& Dehnen}{2005}]{KD05} 
King A.R., Dehnen W., 2005, MNRAS, 357, 275
\bibitem[\protect\citeauthoryear{K{\"o}rding, Falcke \& Markoff}{2002}]{Kording02} 
K{\"o}rding E., Falcke H., Markoff S., 2002, A\&A, 382, L13
\bibitem[\protect\citeauthoryear{Krauss et al.}{2005}]{Krauss05} 
Krauss M., Kilgard R., Garcia M., Roberts T.P., Prestwich A., 2005,
ApJ, 630, 228
\bibitem[\protect\citeauthoryear{Kubota \& Done}{2004}]{KD04} 
Kubota A., Done C., 2004, MNRAS, 353, 980
\bibitem[\protect\citeauthoryear{Kubota et al.}{2001}]{Kubota01} 
Kubota A., Mizuno T., Makishima K., Fukazawa Y., Kotoku J., Ohnishi
T., Tashiro M., 2001, ApJ, 547, L119
\bibitem[\protect\citeauthoryear{Kuntz et al.}{2005}]{Kuntz05} 
Kuntz K.D., Gruendl R.A., Chu Y.-H., Chen C.-H.R., Still M., Mukai K.,
Mushotzky R.F., 2005, ApJ, 620, L31
\bibitem[\protect\citeauthoryear{Lira et al.}{2002}]{Lira02} 
Lira P., Ward M.J., Zezas A., Alonso-Herrero A., Ueno S., 2002, MNRAS,
330, 259
\bibitem[\protect\citeauthoryear{Liu \& Bregman}{2005}]{LiuBreg05}
Liu J.-F., Bregman J.N., 2005, ApJS, 157, 59
\bibitem[\protect\citeauthoryear{Liu, Bregman \& Irwin}{2006}]{LBI06}
Liu J.-F., Bregman J.N., Irwin J., 2006, ApJ, 642, 171
\bibitem[\protect\citeauthoryear{Liu, Bregman \& Seitzer}{2004}]{Liu04} 
Liu J.-F., Bregman J.N., Seitzer P., 2004, ApJ, 602, 249
\bibitem[\protect\citeauthoryear{McClintock \& Remillard}{2006}]{McClRem06}
McClintock J.E., Remillard R.A., 2006, in: ``Compact Stellar X-ray
Sources'', eds. Lewin W.H.G. and van der Klis M., Cambridge University
Press (Cambridge), p. 157
\bibitem[\protect\citeauthoryear{McHardy et al.}{2006}]{McHardy06}
McHardy I.M., Koerding E., Knigge C., Uttley P., Fender R.P., 2006, Nature, 444, 730
\bibitem[\protect\citeauthoryear{Makishima et al.}{2000}]{Maki00} 
Makishima K. et al., 2000, ApJ, 535, 632
\bibitem[\protect\citeauthoryear{Miller et al.}{2003}]{MFMF03} 
Miller J.M., Fabbiano G., Miller M.C., Fabian A.C., 2003, ApJ, 585, L40
\bibitem[\protect\citeauthoryear{Miller, Fabian \& Miller}{2004a}]{MFM04a} 
Miller J.M., Fabian A.C., Miller M.C., 2004, ApJ, 607, 931
\bibitem[\protect\citeauthoryear{Miller, Fabian, \& Miller}{2004b}]{MFM04b} 
Miller J.M., Fabian A.C., Miller M.C., 2004, ApJ, 614, L117
\bibitem[\protect\citeauthoryear{Miller \& Colbert}{2004}]{MC04} 
Miller M.C., Colbert E.J.M., 2004, Int. J. Mod. Phys. D, 13, 1
\bibitem[\protect\citeauthoryear{Miniutti et al.}{2006}]{Mini06}
Miniutti G., Ponti G., Dadina M., Cappi M., Malaguti G., Fabian A.C.,
Gandhi P., 2006, MNRAS, 373, L1
\bibitem[\protect\citeauthoryear{Mizuno et al.}{2007}]{mizuno07} 
Mizuno T., et al., 2007, PASJ, 59, 257
\bibitem[\protect\citeauthoryear{Mucciarelli et al.}{2006}]{Mucci06} 
Mucciarelli P., Casella P., Belloni T., Zampieri L., Ranalli P., 2006,
MNRAS, 365, 1123
\bibitem[\protect\citeauthoryear{Okajima, Ebisawa \& Kawaguchi}{2006}]{Okajima06}
Okajima T., Ebisawa K., Kawaguchi T., 2006, ApJ, 652, L105
\bibitem[\protect\citeauthoryear{Pakull, Grise \& Motch}{2006}]{PGM06} 
Pakull M.W., Gris{\'e} F., Motch C., 2006, in: Populations of High
Energy Sources in Galaxies (Procs. of the 230th Symposium of the IAU),
Eds. E.J.A. Meurs, G. Fabbiano. Cambridge: Cambridge University Press,
pg. 293
\bibitem[\protect\citeauthoryear{Pakull \& Mirioni}{2002}]{PM02}
Pakull M., Mirioni L., 2002, {\tt astro-ph/0202488}
\bibitem[\protect\citeauthoryear{Portegies Zwart et al.}{2004}]{PZ04} 
Portegies Zwart S.F., Baumgardt H., Hut P., Makino J., McMillan
S.L.W., 2004, Nature, 428, 724
\bibitem[\protect\citeauthoryear{Poutanen et al.}{2007}]{Pout07}
Poutanen J., Lipunova G., Fabrika S., Butkevich A.G., Abolmasov P.,
2007, MNRAS, 377, 1187
\bibitem[\protect\citeauthoryear{Ptak \& Colbert}{2004}]{PtakC04}
Ptak, A., Colbert, E.J.M., 2004, ApJ, 606, 29
\bibitem[\protect\citeauthoryear{Rappaport, Podsiadlowski \& Pfahl}{2005}]{Rappa05} 
Rappaport S.A., Podsiadlowski Ph., Pfahl E., 2005, MNRAS, 356, 401
\bibitem[\protect\citeauthoryear{Roberts \& Colbert}{2003}]{RC03} 
Roberts T.P., Colbert E.J.M., 2003, MNRAS, 341, L49
\bibitem[\protect\citeauthoryear{Roberts et al.}{2001}]{Roberts01} 
Roberts T.P., Goad M.R., Ward M.J., Warwick R.S., O'Brien P.T., Lira
P., Hands A.D.P., 2001, MNRAS, 325, L7
\bibitem[\protect\citeauthoryear{Roberts et al.}{2006}]{Roberts06} 
Roberts T.P., Kilgard R.E., Warwick R.S., Goad M.R., Ward M.J., 2006,
MNRAS, 371, 1877
\bibitem[\protect\citeauthoryear{Roberts \& Warwick}{2000}]{RW00} 
Roberts T.P., Warwick R.S., 2000, MNRAS, 315, 98
\bibitem[\protect\citeauthoryear{Roberts et al.}{2004}]{Roberts04} 
Roberts T.P., Warwick R.S., Ward M.J., Goad M.R., 2004, MNRAS, 349, 1193
\bibitem[\protect\citeauthoryear{Roberts et al.}{2005}]{Roberts05} 
Roberts T.P., Warwick R.S., Ward M.J., Goad M.R., Jenkins L.P., 2005,
MNRAS, 357, 1363
\bibitem[\protect\citeauthoryear{Roberts et al.}{2002}]{Roberts02} 
Roberts T.P., Warwick R.S., Ward M.J., Murray S.S., 2002, MNRAS, 337, 677
\bibitem[\protect\citeauthoryear{Soria et al.}{2007}]{Soria07a}
Soria R., Baldi A., Risaliti G., Fabbiano G., King A., La Parola V.,
Zezas A., 2007, {\tt astro-ph/0705.3977}
\bibitem[\protect\citeauthoryear{Soria}{2007}]{Soria07b}
Soria R., 2007, these proceedings
\bibitem[\protect\citeauthoryear{Soria et al.}{2004}]{Soria04}
Soria R., Motch C., Read A.M., Stevens I.R., 2004, A\&A, 423, 955
\bibitem[\protect\citeauthoryear{Stobbart, Roberts \& Warwick}{2004}]{SRW04} 
Stobbart A., Roberts T.P., Warwick R.S., 2004, MNRAS, 351, 1063
\bibitem[\protect\citeauthoryear{Stobbart, Roberts \& Wilms}{2006}]{SRW06} 
Stobbart A., Roberts T.P., Wilms J., 2006, MNRAS, 368, 397
\bibitem[\protect\citeauthoryear{Strohmayer \& Mushotzky}{2003}]{SM03} 
Strohmayer T.E., Mushotzky R.F., 2003, ApJ, 586, L61
\bibitem[\protect\citeauthoryear{Strohmayer et al.}{2007}]{Stroh07} 
Strohmayer T.E., Mushotzky R.F., Winter L., Soria R., Uttley P.,
Cropper M., 2007, ApJ, 660, 580
\bibitem[\protect\citeauthoryear{Swartz et al.}{2004}]{Swartz04} 
Swartz D.A., Ghosh K.K., Tennant A.F., Wu K., 2004, ApJS, 154, 519
\bibitem[\protect\citeauthoryear{Tremaine, Ostriker \& Spitzer}{1975}]{TOS75}
Tremaine S.D., Ostriker J.P., Spitzer L. Jr., 1975 ApJ, 196, 407
\bibitem[\protect\citeauthoryear{Vierdayanti et al.}{2006}]{Vier06}
Vierdayanti K., Mineshige S., Ebisawa K., Kawaguchi, T., 2006, PASJ, 58, 951
\bibitem[\protect\citeauthoryear{Watarai, Mizuno \& Mineshige}{2001}]{Watarai01} 
Watarai K., Mizuno T., Mineshige S., 2001, ApJ, 549, L77
\bibitem[\protect\citeauthoryear{Wolter, Trinchieri \& Colpi}{2006}]{Wolter06}
Wolter A., Trinchieri G., Colpi M., 2006, MNRAS, 373, 1637
\bibitem[\protect\citeauthoryear{Zhang et al.}{2000}]{Zhang00} 
Zhang S.N., Cui W., Chen W., Yao Y., Zhang X., Sun X., Wu X., Xu H.,
2000, Sci, 287, 1239


%
%
\end{thebibliography}


\end{document}